\documentclass[prd,aps,onecolumn,preprintnumbers, showpacs,
nofootinbib,superscriptaddress,notitlepage,floatfix]{revtex4-1}
\usepackage{amsthm,amsmath,amssymb}
\usepackage{mathtools}
\usepackage{mathrsfs}
\usepackage{graphicx}
\usepackage{slashed}
\usepackage{multirow}
\usepackage[colorlinks, citecolor=blue,anchorcolor=blue,menucolor=blue,
linkcolor=blue,filecolor=blue,runcolor=blue,urlcolor=blue,frenchlinks=true]{hyperref}
\usepackage[normalem]{ulem}
\usepackage{xcolor}

\begin{document}

\title{Pion Valence-Quark Generalized Parton Distribution at Physical Pion Mass}

\author{Huey-Wen Lin}
\affiliation{Department of Physics and Astronomy, Michigan State University, East Lansing, MI 48824}
\affiliation{Department of Computational Mathematics,
  Science and Engineering, Michigan State University, East Lansing, MI 48824}

\preprint{MSUHEP-23-011}

\begin{abstract}
We present the first lattice-QCD $x$-dependent pion valence-quark generalized parton distribution (GPD) calculated directly at physical pion mass using the Large-Momentum Effective Theory (LaMET) with next-to-next-to-leading order perturbative matching correction.
We use clover fermions for the valence action on $2+1+1$ flavors of highly improved staggered quarks (HISQ), generated by MILC Collaboration, with lattice spacing $a \approx 0.09$~fm and box size $L \approx 5.5$~fm; the pion two-point measurements number up to $O(10^6)$ with boost momentum 1.73~GeV.
The pion valence distribution is renormalized in hybrid scheme with Wilson-line mass subtraction at large distances in coordinate space, followed by a procedure to match it to the $\overline{\text{MS}}$ scheme.
We focus on the zero-skewness limit, where the GPD has a probability-density interpretation in the longitudinal Bjorken $x$ and the transverse impact-parameter distributions.
We take the integral of our GPD functions to generate leading moment so that we can make comparisons with past lattice-QCD and experimental determinations of the pion form factors and found consistent agreement among them.
We predict the higher GPD moments and reveal $x$-dependent tomography of the pion for the first time using lattice QCD.

\end{abstract}

\maketitle

\section{Introduction}
Pions, as the Nambu-Goldstone bosons created by the chiral symmetry breaking of quantum chromodynamics (QCD), play a crucial role in our understanding of the origin of mass for matter~\cite{Achenbach:2023pba,Arrington:2021biu,Roberts:2021nhw,Aguilar:2019teb}.
Decades of effort have been devoted to pursuing understanding of how quarks and gluons give rise to pions.
Among the various properties of the pions, generalized parton distributions (GPDs), a concept introduced more than two decades ago~\cite{Ji:1996nm,Radyushkin:1997ki}, have been among the most deeply studied both theoretically and experimentally.
(We refer interested readers to the review papers~\cite{Diehl:2003ny,Belitsky:2005qn,Kumericki:2016ehc,Burkert:2022hjz}.)
GPDs retain information on the form factors' dependence on the transfer momentum $Q^2$ as well as the one-dimensional structure in terms of the Bjorken-$x$ parameter in the parton distribution functions (PDFs).
Multiple exclusive processes, such as deeply virtual Compton scattering (DVCS) and deeply virtual meson production (DVMP), provide experimental access to the GPDs, allowing us to map out the tomography of hadrons~\cite{Burkardt:2000za}. 
DVCS has been extensively studied during the last two decades;
the process is relatively simple in its interpretation, since the only nonperturbative object entering its amplitudes, which are referred to as Compton form factors (CFFs), are the GPDs.
They are the main source of information on GPDs.
We refer readers to Ref.~\cite{Kumericki:2016ehc} for phenomenological context. 
Efforts have been made to extract information experimentally about the
internal structure of pion, ranging from its electromagnetic form factor through pion-electron scattering~\cite{NA7:1986vav} to its PDFs through the Drell-Yan (DY) process~\cite{Conway:1989fs,NA10:1987azm,NA10:1985ibr}. 
As a result, there has been much progress made toward exploring the global analysis of pion PDFs in recent years~\cite{Barry:2018ort,Novikov:2020snp,Barry:2021osv}, but not much effort has been made toward pion GPDs. 
There are many ongoing and planned experimental efforts to further our knowledge of pion structure; for example 
the JLab 12-GeV program will provide precision data relating to the pion and kaon form factors up to $Q^2 \approx 10\mbox{ GeV}^2$ and $5\mbox{ GeV}^2$, respectively.
They will also measure the pion and kaon structure functions at $x > 0.5$ through the Sullivan process~\cite{Dudek:2012vr}. 
AMBER Collaboration at CERN can play a crucial role due to their unique capability delivering pion and kaon DY measurements in the centre-of-mass (CoM) energy range 10--20~GeV. 
A future facility, the Electron-Ion Collider (EIC) at Brookhaven National Laboratory, will provide information on pion and kaon structure over a large and tunable CoM energy range 20--140~GeV via Sullivan-process measurements~\cite{Accardi:2012qut,Arrington:2021biu,Aguilar:2019teb,AbdulKhalek:2021gbh,Achenbach:2023pba,Burkert:2022hjz}. 
An electron-ion collider being discussed in China (EicC)~\cite{Anderle:2021wcy,Chavez:2021koz} would have a similar CoM energy range to AMBER and neatly fill a gap between JLab~12 and the EIC.

Lattice quantum chromodynamics (LQCD) is an ideal nonperturbative theoretical method, allowing us to study pion structure via the QCD path integral with full systematic control.
For a long while, LQCD was limited to low-order moments of PDFs and distribution amplitude, form factors and generalized form factors of the pion.
Since 2013, numerous calculations of $x$-dependent hadron structure in lattice QCD have emerged since the proposal of Large-Momentum Effective Theory (LaMET)~\cite{Ji:2013dva,Ji:2014gla,Ji:2017rah}.
Many lattice works have been done on nucleon and meson PDFs and GPDs based on the quasi-PDF approach~\cite{Lin:2013yra,Lin:2014zya,Chen:2016utp,Lin:2017ani,Alexandrou:2015rja,Alexandrou:2016jqi,Alexandrou:2017huk,Chen:2017mzz,Zhang:2018diq,Alexandrou:2018pbm,Chen:2018xof,Chen:2018fwa,Alexandrou:2018eet,Lin:2018qky,Fan:2018dxu,Liu:2018hxv,Wang:2019tgg,Lin:2019ocg,Chen:2019lcm,Lin:2020reh,Chai:2020nxw,Bhattacharya:2020cen,Lin:2020ssv,Zhang:2020dkn,Li:2020xml,Fan:2020nzz,Gao:2020ito,Lin:2020fsj,Zhang:2020rsx,Alexandrou:2020qtt,Lin:2020rxa,Gao:2021hxl,Lin:2020rut,Alexandrou:2020zbe,Lin:2021brq,Bhattacharya:2022aob}.
Alternative approaches to access $x$-dependent structure in lattice QCD are also proliferating; for example,
the Compton-amplitude approach (or ``OPE without OPE'')~\cite{Aglietti:1998ur,Martinelli:1998hz,Dawson:1997ic,Capitani:1998fe,Capitani:1999fm,Ji:2001wha,Detmold:2005gg,Braun:2007wv,Chambers:2017dov,Detmold:2018kwu,QCDSF-UKQCD-CSSM:2020tbz,Horsley:2020ltc,Detmold:2021uru},
the ``hadronic-tensor approach''~\cite{Liu:1993cv,Liu:1998um,Liu:1999ak,Liu:2016djw,Liu:2017lpe,Liu:2020okp},
the ``current-current correlator''~\cite{Braun:2007wv,Ma:2017pxb,Bali:2017gfr,Bali:2018spj,Joo:2020spy,Gao:2020ito,Sufian:2019bol,Sufian:2020vzb}
and the pseudo-PDF approach~\cite{Radyushkin:2017cyf,Balitsky:2019krf,Orginos:2017kos,Karpie:2017bzm,Karpie:2018zaz,Karpie:2019eiq,Joo:2019jct,Joo:2019bzr,Radyushkin:2018cvn,Zhang:2018ggy,Izubuchi:2018srq,Joo:2020spy,Bhat:2020ktg,Fan:2020cpa,Sufian:2020wcv,Karthik:2021qwz,HadStruc:2021wmh,Fan:2021bcr,HadStruc:2022yaw,Salas-Chavira:2021wui,Fan:2022kcb}.
A few works have started to include lattice-QCD systematics, such as finite-volume effects~\cite{Lin:2019ocg,Sufian:2020vzb} and lattice-spacing dependence for quark~\cite{Lin:2020fsj,Alexandrou:2020qtt,Karpie:2021pap,Zhang:2020gaj,Lin:2020ssv,Gao:2022iex}
and gluon~\cite{Fan:2022kcb,Salas-Chavira:2021wui,Fan:2021bcr} distributions, in their $x$-dependent structure calculations.
Most lattice calculations of PDFs use next-to-leading-order (NLO) matching~\cite{Xiong:2013bka,Ma:2014jla,Ji:2017rah, Ji:2020ect}, but recently some lattice calculations of the valence pion PDF~\cite{Gao:2021dbh} have incorporated NNLO matching~\cite{Chen:2020ody,Li:2020xml}.

Less progress has been made toward $x$-dependent GPD studies by contrast.
The first lattice study of GPDs was made for pions~\cite{Chen:2019lcm} in 2019 with largest boost momentum 1.7~GeV using 310-MeV pion mass.
By 2020, ETMC~\cite{Alexandrou:2020zbe} reported results at a single $Q^2$ for both unpolarized and polarized GPDs with a 260-MeV pion with skewness of 0 and 0.3 with a single transfer momentum of 1.67~GeV.
In the same and following year, MSULat~\cite{Lin:2020rxa,Lin:2021brq} reported on the unpolarized and polarized zero-skewness GPDs with 135-MeV pion at multiple $Q^2$.
One can actually take the moments of MSULat's $x$-dependent GPDs and compare with lattice results using a traditional moment calculation using a local operator via the OPE method, such as electromagnetic and axial form factors (with $n=1$), and generalized form factors ($n=2$).
With multiple $Q^2$ values in the zero-skewness limit, the Fourier transform of the non--spin-flip GPD $H(x,\xi=0,-Q^2)$ yields the impact-parameter--dependent distribution, which cannot be directly accessed via experimental measurements.
Progress so far has been done using Breit frame, where the initial and final momenta of the hadron differ by half the transfer momentum.
Recent work on asymmetric-momentum setups for GPDs has been demonstrated by the ETM and BNL/ANL groups~\cite{Bhattacharya:2022aob} and can help reduce the computational cost of the lattice calculation.
Pion structure, such as form factors, can be more sensitive to the pion mass used in the lattice calculation; therefore, it is important to study pion structure directly at the physical pion mass.

In this work, we report the first study at physical pion mass of the pion unpolarized valence-quark GPD, using LaMET method with lattice spacing 0.09~fm in the Breit frame.
The remainder of this paper is organized as follows:
In Sec.~\ref{sec:cal-details}, we discuss the lattice setup and the procedure for how the lattice two- and three-point correlators are calculated and analyzed to extract the ground-state matrix elements.
These matrix elements are then renormalized using hybrid renormalization scheme and the valence-quark GPDs are extracted by fitting the matrix elements with NNLO kernels in Sec.~\ref{sec:results}.
We compare our results in certain limits that have prior LQCD calculations for comparison: our valence-quark GPD at $Q^2$ result is consistent with prior lattice physical-pion-mass calculate the quasi-GPD distribution and matched them to the lightcone, and the first moment of our GPD is in good agreement with the prior lattice pion form factors.
We predict the transfer-momentum dependence of the pion GPDs, higher moments of the generalized form factors that have not been calculated on the lattice, and the tomography of the pion.
Conclusions and future outlook can be found in Sec.~\ref{sec:summary}.

\section{Calculation Setup } 
\label{sec:cal-details}

The unpolarized valence-quark GPD of the pion on the lightcone is defined as
\begin{equation}\label{GPD}
H_q^\pi(x,\xi,t,\mu) = \int\frac{d\eta^-}{4\pi} e^{-i x \eta^- P^+} 
 \times \left\langle \pi\left(P+\frac{\Delta}{2}\right)\left|{\bar q}\left(\frac{\eta^-}{2}\right)\gamma^+ \Gamma\left(\frac{\eta^-}{2}, -\frac{\eta^-}{2}\right)q\left(-\frac{\eta^-}{2}\right)\right|\pi\left(P-\frac{\Delta}{2}\right)\right\rangle,
\end{equation}
where
Bjorken-$x$ is the momentum fraction $x\in[-1,1]$, 
$\mu$ is the renormalization scale in the $\overline{\text{MS}}$ scheme,
momentum $P^\mu=(P^0,0,0,P^z)$, where the pion momentum of the initial and final states are $P \mp \Delta/2$ with $\Delta$ the momentum transfer,
and the variables $t=\Delta^2$ and $\xi=-\frac{\Delta^+}{2P^+}$. 
$\bar q$ and  $q$ represent the antiquark and quark fields, $\eta^{\pm} = (\eta^0 \pm \eta^3)/\sqrt 2$,
and
the gauge link 
$\Gamma(\eta_2^-,\eta_1^-) = \exp\left(-ig \int_{\eta_1^-}^{\eta_2^-}d\eta^- A^+(\eta^-) \right)$
ensures gauge invariance of the quark bilinear operator.
In the forward limit ($\Delta^{\mu} \to 0$), it reduces to the PDF.

On the lattice, one can compute ground-state matrix elements of a pion with a finite-$P^z$ boost in the Breit frame,
\begin{equation}
\label{eq:qlat}
\tilde{h}_\text{lat}(z,P^z,t,a) =  \frac{1}{2P^0} \times {} 
  \left\langle \pi^{+}\left(\vec{P}+\frac{\vec{\Delta}}{2}\right) \right|
    \bar{q}(z) \Gamma \left( \prod_n U_z(n\hat{z})\right)\tau_{3} q(0)
  \left| \pi^{+}\left(\vec{P}-\frac{\vec{\Delta}}{2}\right) \right\rangle,
\end{equation}
where $U_z$ is a discrete gauge link in the $z$ direction, $\vec{P}=\{0,0,P^z\}$ is the 3-momentum of the pion, $\Gamma=\gamma^t$ and $\vec{\Delta}$ is the momentum transfer between initial and final pion. 
The matrix elements are renormalized and then contribute to the quark quasi-GPD via
\begin{equation}
\label{eq:qGPD}
\tilde{H}_q^\pi(x,\xi,t,P^z,\tilde\mu) = \int\frac{d z P^z}{2\pi} e^{i x P^z z}\tilde h^R(z, P^z, \xi, t, \tilde\mu)
\end{equation}
using the renormalized matrix element $\tilde h^R$.
The lightcone GPD in the $\overline{\text{MS}}$ scheme at scale $\mu$ is then convolved with a perturbative hard matching kernel, up to power corrections that are suppressed by the pion momentum~\cite{Liu:2019urm}
\begin{equation}
\label{eq:matching}
\tilde{H}_{u-d,R}^\pi(x,\xi,t,P^z) 
= \int
{dy\over |y|}C\left({x\over y},\frac{\xi}{y},
\frac{yP^z}{\mu},
\right)H_{u-d}^\pi(y,\xi,t,\mu) 
+\mathcal{O}\left(
{ \Lambda_\text{QCD}^2\over {x^2 P_z^2}}, {\Lambda_\text{QCD}^2\over {(1-x)^2P_z^2}}\right).
\end{equation}
In the zero skewness limit $\xi=0$, the matching kernel $C$ is the same as the matching kernel for the PDF, as documented in Refs.~\cite{Chen:2018xof,Liu:2018uuj}.

This calculation is carried out using $N_f = 2+1+1$ highly improved staggered quarks (HISQ)~\cite{Follana:2006rc}, generated by the MILC Collaboration~\cite{Bazavov:2012xda}, with lattice spacing $a\approx 0.09$~fm at physical pion mass.
Five steps of hypercubic (HYP) smearing~\cite{Hasenfratz:2001hp} are applied to the gauge links to reduce short-distance noise.
We use Wilson-clover fermions in the valence sector and tune the valence-quark masses to reproduce the lightest light and strange sea pseudoscalar meson masses.
A similar setup is used by PNDME collaboration~\cite{Mondal:2020cmt,Park:2020axe,Jang:2019jkn,Jang:2019vkm,Gupta:2018lvp,Lin:2018obj,Gupta:2018qil,Gupta:2017dwj,Bhattacharya:2015wna,Bhattacharya:2015esa,Bhattacharya:2013ehc} with local operators, such as isovector and flavor-diagonal charges, form factors and moments;
the results from this mixed-action setup are consistent with the same physical quantities calculated using different fermion actions~\cite{Kronfeld:2019nfb,Lin:2017snn,Lin:2020rut,Lin:2022nnj,FlavourLatticeAveragingGroup:2019iem,FlavourLatticeAveragingGroupFLAG:2021npn}.
BNL/ANL Collaboration has also been using a mixed-action setup for their pion valence-quark PDFs calculations~\cite{Gao:2021dbh,Gao:2022iex}.
We carefully monitor the measurements from each configuration to rule out any possibility of exceptional configurations;
we have not observed any here nor in the previous works mentioned earlier.

To calculate the GPD matrix elements at nonzero momentum transfer, we first calculate the matrix element $\langle \chi_\pi (\vec{P}_f) | O^\mu | \chi_\pi (\vec{P}_i) \rangle$, where $\chi_\pi$ is the pion interpolating field,
$O_\mu=\overline{\psi}\gamma_\mu W(z) \psi$ is the LaMET Wilson-line displacement operator with $\psi$ being the quark field, and $\vec{P}_{\{i,f\}}$ the initial and final nucleon momenta. 
We use 1960 configurations to help with statistical noise at boost momentum $P_z=\left|\frac{\vec{P}_i+\vec{P}_f}{2}\right| = \left|\frac{2\pi}{L}\{0,0,8\}a^{-1}\right|$.
We vary the spatial momentum transfer $\vec{q}=\vec{P}_f-\vec{P}_i=\frac{2\pi}{L}\{n_x,n_y,0\}a^{-1}$ with integer $n_{x,y}$ and $n_x^2+n_y^2 \in \{0,4,8,16,20\}$
with four-momentum transfer squared { $t=Q^2=-q_\mu q^\mu \in \{{0, 0.19, 0.39, 0.77, 0.97}\} $}~GeV$^2$ 
using periodic boundary conditions.
We use source-sink separations of \{6,7,8,9,10\} lattice units with
high-statistics measurements of $\{\text{752,640}, \text{1,003,520}, \text{1,003,520}, \text{1,505,280}, \text{1,505,280}\}$, respectively.
We extract pion ground-state matrix elements using two-state simultaneous fits about which more details can be found in our previous work~\cite{Lin:2018obj,Lin:2018qky,Lin:2020rxa,Lin:2021brq}.

\begin{figure}[tb]
\includegraphics[width=0.42\textwidth]{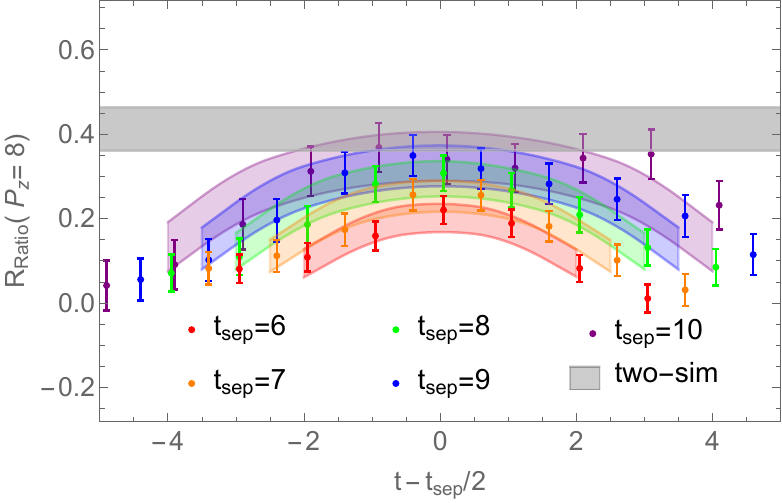}
\caption{
An example of the bare ground-state matrix-element determination using two-state fits to multiple source-sink separations, along with the ratio plot and reconstructed fit using $\vec{P_f}=\{1,0,8\}\frac{2\pi}{L}$ and $\vec{P_i}=\{-1,0,8\}\frac{2\pi}{L}$.  The colored bands are the
reconstructed ratios from the fits to source-sink separation
$t_{\rm sep} \approx [0.72,0.9]$~fm, respectively, and the gray band shows the
ground-state matrix elements from the “two-simRR” fits.
\label{fig:ratio}}
\end{figure}

Figure~\ref{fig:ratio} shows an example of the ground-state matrix elements (shown as the gray band) from the
simultaneous two-state fitted results using source-sink separation of 
$t_\text{sep} \in [6,10]$ in lattice units,
at $\vec{P_f}=\{1,0,8\}\frac{2\pi}{L}$ and $\vec{P_i}=\{-1,0,8\}\frac{2\pi}{L}$.
The signal in the three-point correlators decreases as the source-sink separation increases;
also as expected, the ground state for each $t_\text{sep}$ approaches the simultaneous-fit ground-state matrix element, shown as the gray band in the plot.
 The spectral decomposition predicts that the data for all three quantities is
symmetric about $t = t_\text{sep} /2$ in Breit frame; at large $t_\text{sep}$ such symmetry breaks slightly due to the statistical fluctuations. The ground-state matrix elements are consistent when one removes smaller $t_\text{sep}$ from the analysis.

\begin{figure}[tb]
\includegraphics[width=0.42\textwidth]{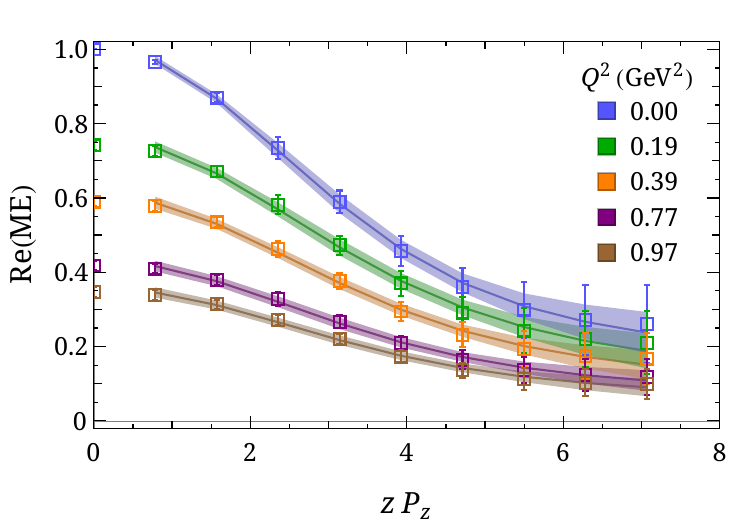}
\includegraphics[width=0.42\textwidth]{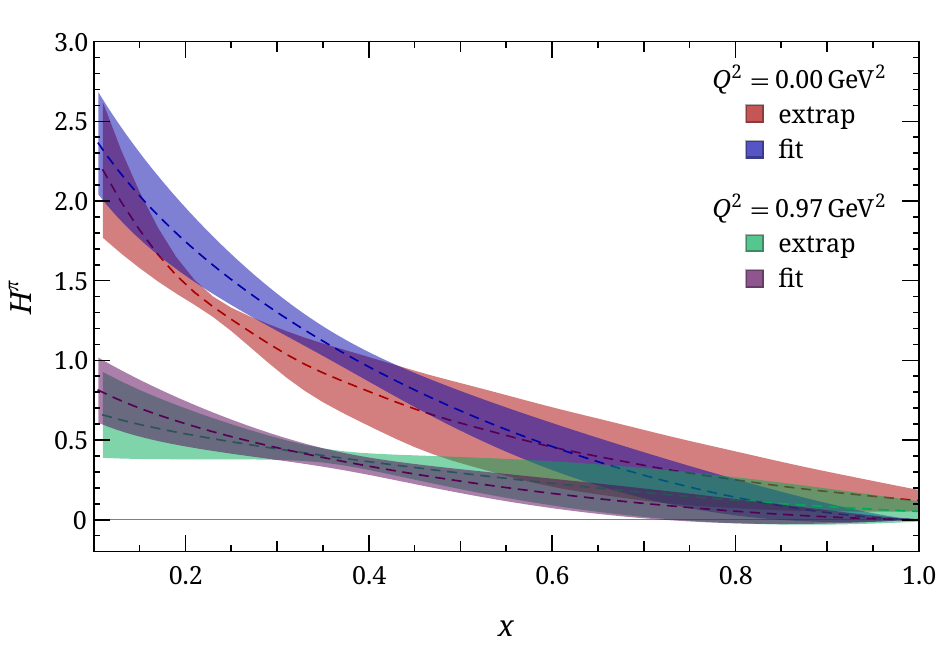}
\caption{
(Left) The hybrid renormalized Wilson-line length displacement $z$-dependence of matrix elements as a function of dimensionless parameter $zP_z$ at various momentum transfer $Q^2\in\{0, 0.97\}\text{ GeV}^2$.
The corresponding bands are the reproducing matrix elements from the fit to the matrix elements using Eq.~\ref{eq:matching}. 
(Right) Comparison of two methods, fit-matching ``fit'' and direct matching with extrapolation ``extrap'', to extract the lightcone distributions for valence-quark pion GPD at $Q^2=0$ and $0.97 \text{ GeV}^2$.
The two methods are consistent over most $x$ values within statistical errors.
There is some tension near $x=0.25$ and $x>0.9$ for $H^\pi (x,Q^2=0)$ GPD, but they are still consistent within two sigma.
\label{fig:H-MEs}}
\end{figure}

\section{Results and Discussion}
\label{sec:results}

We use the hybrid renormalization scheme~\cite{Ji:2020brr} with NNLO perturbative matching correction~\cite{Chen:2020ody,Li:2020xml}.
The exact procedure has been applied and detailed in prior valence-quark PDF studies by BNL/ANL group~\cite{Gao:2021dbh,Gao:2022iex}. 
This renormalization scheme is a combination of the ratio scheme at short distances up to Wilson-line length $z_s$ and an explicit subtraction of the self-energy divergence of the Wilson line at large distances.
The parameter $z_s$ that separates the nonperturbative and perturbative regions must be much larger than the lattice spacing to avoid discretization effects but not so large as to necessitate higher-twist terms in the OPE.
For our calculations, we choose $z_s=0.27$~fm.
The hybrid normalized matrix elements as functions of $zP_z$ can be found on the left-hand side of Fig.~\ref{fig:H-MEs}.
We also normalize all the matrix elements by the bare ground-state matrix element at $z=0$ and $Q^2=0$ (vector charge normalized to 1) to improve the signal. 
The real matrix elements decrease quickly to zero as a function $z$ due to the large boost momentum used in this calculation.
The signal-to-noise ratio also increases quickly as one increases the Wilson-line displacement.
The central values of matrix elements also drop quickly as $Q^2$ increases initially but slow down at the larger $Q^2$;
this is expected, since similar behavior observed in the $Q^2$-dependence of the pion form factors.

The valence-quark GPD $H^\pi(x,Q^2)$ is then extracted using Eq.~\ref{eq:matching} with NNLO matching kernel $C$ from Refs.~\cite{Chen:2020ody,Li:2020xml}. 
We compare two methods for performing the matching.
First, we extrapolate the matrix elements to large $\lambda = zP_z$ using a form with the exponential and power decay, $d_1 e^{-d_2 |z|}/|\lambda|^{d_3}$, as used in prior pion-PDF studies~\cite{Gao:2021dbh,Gao:2022iex}.
The fit parameters $d_1$, $d_2$ and $d_3$ vary with $Q^2$ and are fit to the data subset at large $zP_z > 3.9$.
We then Fourier transform the matrix elements into the quasi-distribution and apply matching according to Eqs.~\ref{eq:qGPD} and \ref{eq:matching}.
Alternatively, we can adopt a phenomenology-inspired functional form to describe the $x$ dependence of the lightcone GPD: 
\begin{equation} 
H^\pi(x,Q^2; m,n,c ) =
  \frac{x^m (1-x)^n (1+c\sqrt{x})}
  {B(m+1,n+1) + c B\left(m+\frac{3}{2},n+1\right)},
\label{eq:PDFfitFunc}
\end{equation}
where the parameters $m$, $n$ and $c$ are $Q^2$-dependent, and
the beta function $B(m+1,n+1)=\int_0^1 dx\,x^m (1-x)^n$ is used to normalize to unity at $Q^2=0$.
Note that this form is also commonly used by the PDF global-fit community, for example, in the pion PDF fit by JAM~\cite{Barry:2018ort,Cao:2021aci}.
This phenomenological form has also been widely used for $x$-dependent valence-quark pion PDFs, such as in Refs.~\cite{Sufian:2019bol,Joo:2019bzr,Izubuchi:2019lyk,Lin:2020ssv,Sufian:2020vzb,Gao:2022iex}.
We perform the phenomenological fit to our physical pion-mass ensemble using the matrix elements with Wilson-line displacement $z\in [0.09, 0.81]$~fm for each transfer momentum.
We choose the $\overline{\text{MS}}$ renormalization scale to be $\mu = 2$~GeV.
We find the goodness of fit to be around 1 with either $c$ as a free parameter or $c=0$.  
As in the previous lattice valence-quark pion PDF study, the final distributions are consistent with either choice, but the former case results in the distribution having larger error bands due to the additional free parameter.
For the remainder of this work, we focus on the $c=0$ results, since this is the first lattice work on the pion GPD at physical pion mass.
Further study of the systematics due to fit-form choices, as well as other lattice artifacts (lattice discretization, large momentum dependent, etc.), should be explored when more computational resources are available.
The reconstructions of the matrix elements using the fit parameters are shown as the bands on the left-hand side of Fig.~\ref{fig:H-MEs};
they pass through the input data nicely.
Varying $z_\text{max}$ results in small changes to the goodness-of-fit and $x$ dependence of the distributions. 
The two methods of extracting the GPD are consistent for all $Q^2$ values studied here.
We show two selected values of $Q^2$ results on the right-hand side of Fig.~\ref{fig:H-MEs}, labeled as ``extrap'' and ``fit'' for the former and latter methods, respectively.
Both results are consistent in the intermediate-$x$ region within statistical errors.
Some small disagreement appears for $x$ around 0.25 and at large $x$ of $H^\pi(x,Q^2=0)$ but still within two standard deviations;
this may change as the statistics increase in future updates. 
For the remainder of the manuscript, we focus on the results from the fitting method.

\begin{figure*}[tb]
\includegraphics[width=0.45\textwidth]{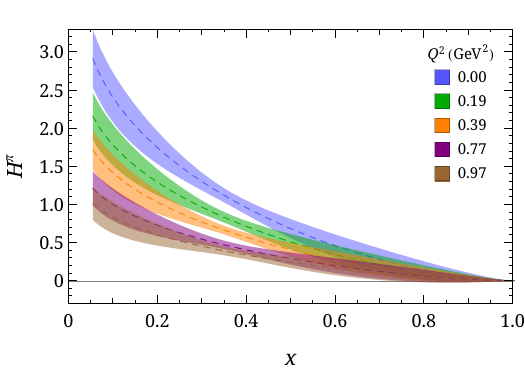}
\includegraphics[width=0.45\textwidth]{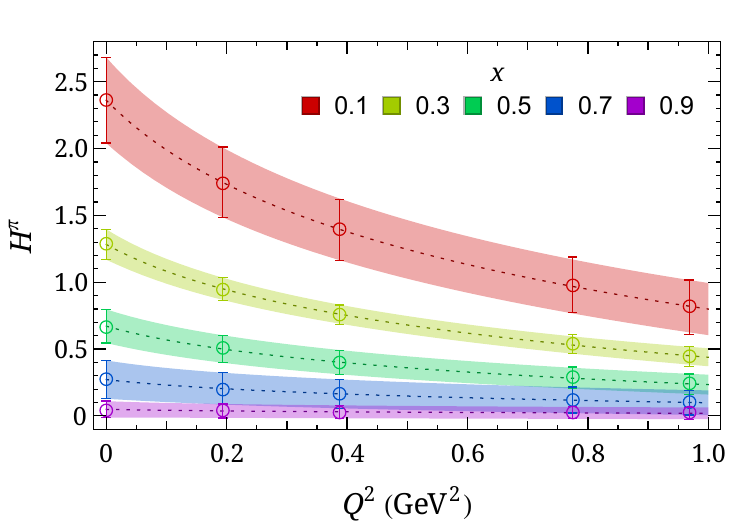}
\caption{
(left)
Pion valence-quark GPD as a function of Bjorken-$x$ with five values of transfer momenta studied in this work.
(right)
Pion valence-quark GPD as a function of transfer momenta at selected Bjorken-$x$ indicated in the bands.
The $z$-expansion is used to interpolate between the five transfer momenta.
\label{fig:pionGPD}}
\end{figure*}

Figure~\ref{fig:pionGPD} shows our final result for the pion valence-quark GPD.
On the left-hand side of Fig.~\ref{fig:pionGPD}, the GPD at zero transfer momentum as a function of Bjorken-$x$ (that is, the PDF) is shown, as well as the distribution at various transfer momenta in the zero-skewness limit.
As the transfer momentum increases, the distribution decreases, as expected.
On the right-hand side of Fig.~\ref{fig:pionGPD}, we show the GPD at selected $x$ values as functions of transfer momentum $Q^2$ (in units of GeV$^2$).
The GPD is fitted with a $z$-expansion up to 3 parameters to interpolate the $Q^2$ dependence.
The $Q^2$ dependence of the GPD can help us investigate the tomography of the pion. 
Unfortunately, there are few published lattice pion GPDs at physical pion mass to make comparisons. 
In first lattice GPD work~\cite{Chen:2019lcm}, the pion valence-quark GPD at zero skewness was calculated using clover valence fermions on an ensemble of gauge configurations
with lattice spacing $a \approx 0.12$~fm, box size $L \approx 3$~fm and pion mass $m_\pi \approx 310$~MeV.
Our results are consistent with the heavier quark mass but with the systematics removed or improved for being at the physical pion mass and smaller lattice spacing. 
We can compare our results at zero transfer momentum with the PDF results of BNL/ANL group, also done at physical pion mass but at a finer lattice spacing of 0.076~fm~\cite{Gao:2021dbh,Gao:2022iex};
they are consistent within statistical errors.

\begin{figure*}[t]
\includegraphics[width=0.45\textwidth]{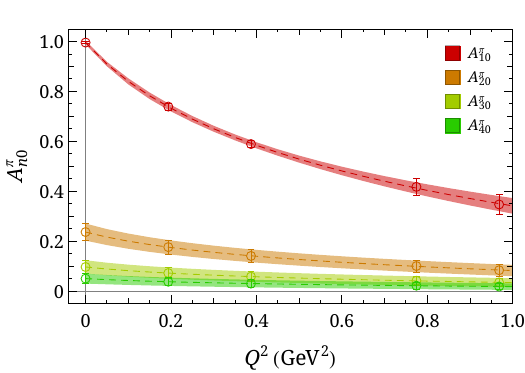}
 \includegraphics[width=0.45\textwidth]{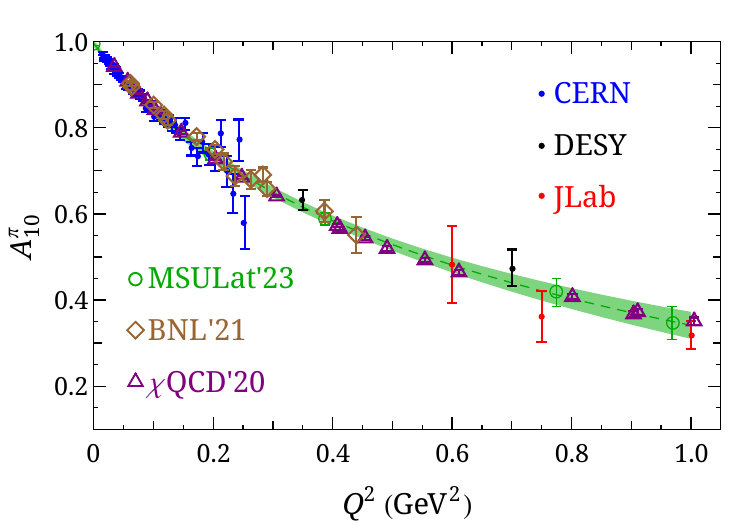}
\caption{\label{Fig:GPDmoments}
(left) Lowest four unpolarized pion GFFs $A_{n0}$ with $n\in[1,4]$ obtained from taking the moment integral using the pion GPD function obtained from this work.
(right)
Selected pion form factor $F_\pi(Q^2)$ at the physical pion mass flavours of light quarks from different lattice groups (labeled ``$\chi$QCD'20''~\cite{Wang:2020nbf},  ``BNL'21''~\cite{Gao:2021xsm}), together with the result obtained in this work (labeled ``MSULat'23'') and experimental data \cite{NA7:1986vav}.
We found that this leading moments of our pion GPD is in agreement with prior lattice works and existing experimental data~\cite{JeffersonLab:2008jve,JeffersonLab:2008gyl,Horn:2007ug,JeffersonLabFpi-2:2006ysh,JeffersonLabFpi:2000nlc}.
}
\end{figure*}

The zero-skewness limit of the pion valence-quark GPD is related to the Mellin moments by taking the $x$-moments~\cite{Ji:1998pc,Hagler:2009ni}
\begin{equation}
\label{eq:GFFs}
\int_{-1}^{+1}\!\!dx \, x^n \, {H^\pi}(x, \xi, Q^2)=  
{A}_{n+1,i}^\pi(Q^2),
\end{equation}
where ${A}_{ni}^\pi(Q^2)$ are the generalized form factors (GFFs). Our results for $n \in [1,4]$ can be found in the left-hand-side of Fig.~\ref{Fig:GPDmoments}. The GFF allows us to make comparison with past LQCD calculation and provide possible constraints for global analysis. 
In the past, there have been some works done to use the operator product expansion (OPE) to directly calculate Mellin moments of the GPDs on the lattice using local matrix elements.
Generally, the OPE yields
\begin{widetext}
\begin{multline}
 \left\langle \pi^+(P+\Delta/2) \left| \bar{u}(0) \gamma^{\{ \mu} iD^{\mu_1} iD^{\mu_2}\dots iD^{\mu_n\}} u(0) \right| \pi+(P-\Delta/2) \right\rangle \\
 = 2P^{\{ \mu} P^{\mu_1}\dots P^{\mu_n \}} A_{n+1,0}(\Delta^2) + 2 \;\sum_{\mathclap{i=1,\text{ odd}}}^n\; \Delta^{\{ \mu}
\Delta^{\mu_1}\ldots \Delta^{\mu_i} P^{\mu_{i+1}} \ldots P^{\mu_n} A_{n+1,i+1}(\Delta^2)\,.\label{eq:pion_gff}
\end{multline}
\end{widetext}
Note that when $n=0$ in Eq.~\ref{eq:pion_gff}, it yields the pion electromagnetic form factor $\left\langle\pi(p')\left| \mathcal{O}^{\mu} \right|\pi(p)\right\rangle = 2\, P^{\mu} F_{\pi}(\Delta^2)$, where $A_{1,0} = F_{\pi}$.
For this case, there is a long history of lattice-QCD calculations using the vector-current operator, and there have also been few attempts to calculate the $A_{2,0}$ moments.
Such calculations will provide independent cross-checks of the results obtained directly from LaMET by taking the integral over $x$ to access the moments.

The elastic electromagnetic form factors of the charged pion are the most studied pion structure in lattice QCD~\cite{Martinelli:1987bh,Draper:1988bp,Bonnet:2004fr,Brommel:2006ww,Frezzotti:2008dr,Aoki:2009qn,Nguyen:2011ek,Chambers:2017tuf,Koponen:2017fvm,Alexandrou:2017blh,Feng:2019geu,Wang:2020nbf,Gao:2021xsm}. 
In recent years, these include work directly calculated at physical pion mass~\cite{Koponen:2015tkr,Alexandrou:2017blh,Feng:2019geu,Wang:2020nbf,Gao:2021xsm}. 
In Fig.~\ref{Fig:GPDmoments}, we compare our pion form factors, obtained from taking the moments of the pion GPD function, with two recent lattice calculations using vector-current operator.
$\chi$QCD Collaboration used overlap fermions on seven ensembles of 2+1-flavor domain-wall configurations, including multiple lattice spacings $a \in [0.083,0.195]$~fm to remove lattice discretization effects, and pion masses ranging from 139 to 340~MeV.
Their statistics range from 9,600 to 485,376 measurements~\cite{Wang:2020nbf}. 
BNL used clover on $N_f=2+1$ HISQ lattice at a single lattice spacing $a = 0.076$~fm at physical pion mass with 1750 and 35,000 exact and sloppy inversions, respectively~\cite{Gao:2021xsm};
the work also includes two additional ensembles at smaller lattice spacing with 300-MeV pion to quantify the systematics.
They found the pion radius to be consistent at heavy pion mass between the 0.04 and 0.06~fm lattice-spacing ensembles but sensitive to the pion mass. 
We select their results at physical pion mass on the smallest lattice-spacing to compare to our integral over the GPD. 
We can see that there is a very nice agreement among all the lattice results, even though the lattice spacing varies among the three calculations.
We also compare the pion form factor with those extracted from experiments~\cite{JeffersonLab:2008jve,JeffersonLab:2008gyl,Horn:2007ug,JeffersonLabFpi-2:2006ysh,JeffersonLabFpi:2000nlc} and find good agreement;
the lattice data are more precise than the available experimental data in certain transfer-momentum regions.

There have been only a couple of dynamical calculations of $A_{2,0}^\pi$ using heavy pion masses more than a decade ago, reported in conference proceedings. 
In 2005, UKQCD/QCDSF~\cite{Brommel:2005ee} used two flavors of dynamical fermion
with a very heavy pion mass of 1090~MeV and found $A_{20}^\pi(t = 0) = 0.315(8)$. 
In 2013, RQCD also used two-flavor clover dynamical fermions with pion masses ranging from 150 to 491 MeV~\cite{Bali:2013gya};
unfortunately, they found $A_{20}^\pi(t)$ to have a wide range distribution and the data at the near-physical pion mass has uncertainty too large for a direct comparison.
However, overall, our moment-integral from GPD functions lies within the range of RQCD moment results obtained using OPE method. 
Unfortunately, there have been no $n \geq 3 $ GFF moment calculations from OPE methods.
Our $A_{\{3,4\}0}^\pi$ results, shown in Fig.~\ref{Fig:GPDmoments}, are the only predictions from LQCD.

\begin{figure*}[tb]
\includegraphics[width=0.6\textwidth]{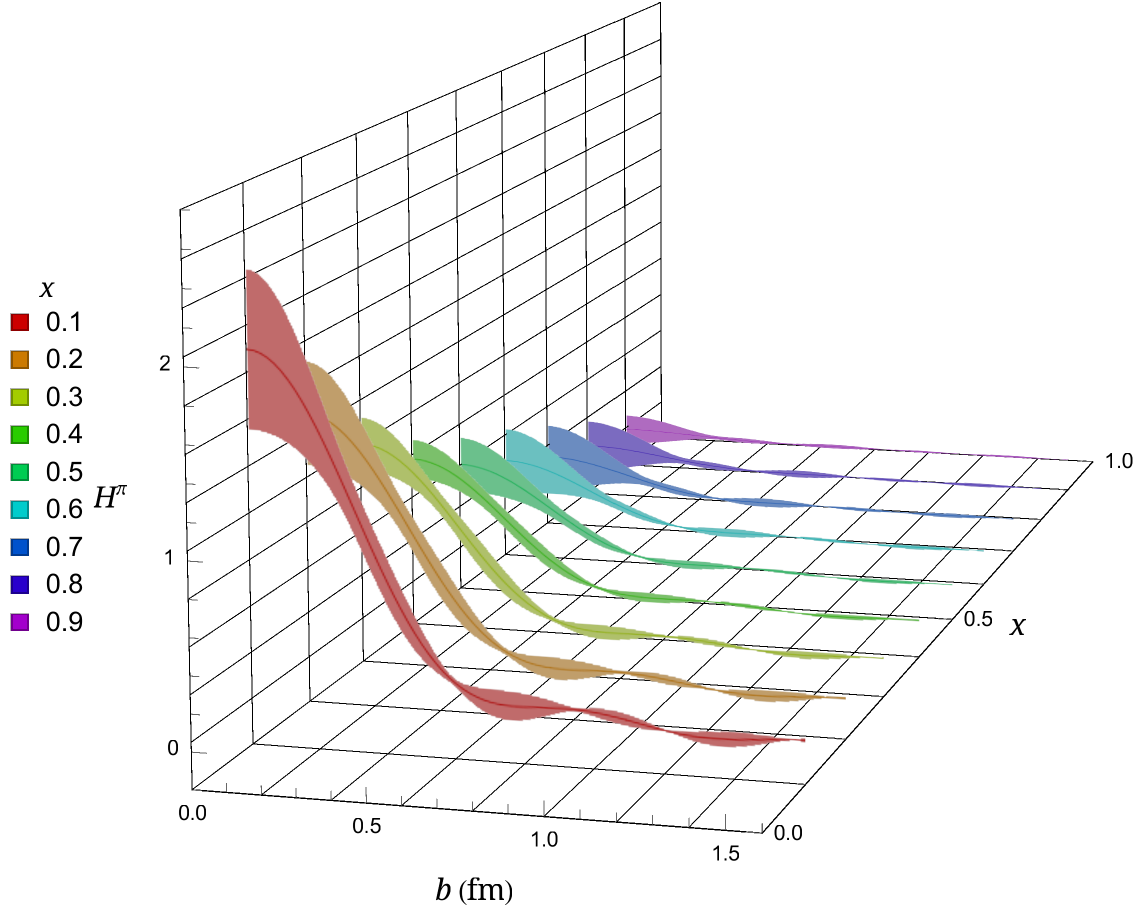}
\includegraphics[width=0.35\textwidth]{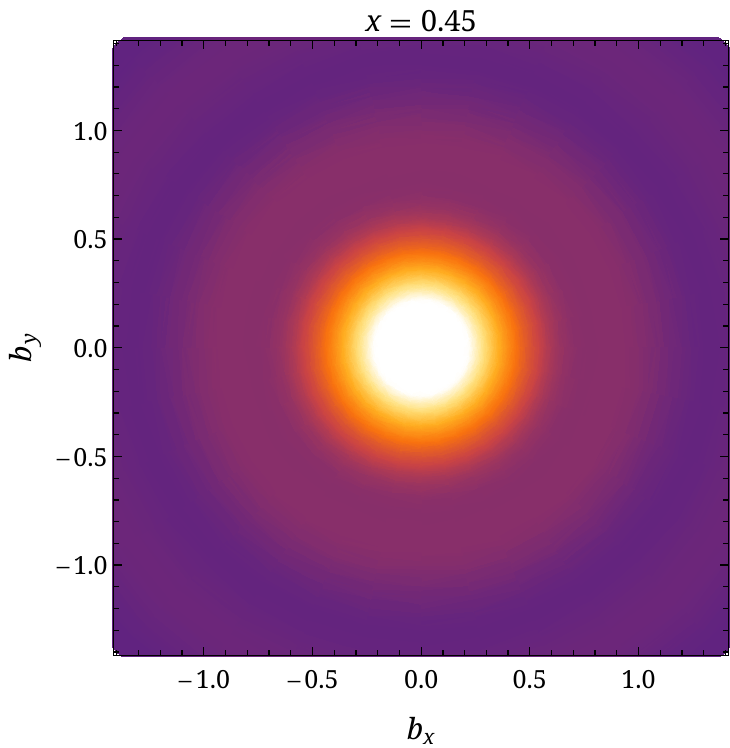}
\caption{
(left)
The valence-quark impact-parameter--dependent distribution of pion as a function of $b$ for selected $x$ and
(right)
an example of the two-dimensional distribution at $x=0.45$.
All distances are in units of fm.
\label{fig:tomography}}
\end{figure*}

Taking the
lattice calculations of the pion's valence quark GPD, $H_\pi(x,\xi=0,Q^2)$, we can then Fourier transform of this GPD to learn about the impact-parameter--dependent distribution, $\mathsf{q}(x,b)$~\cite{Burkardt:2002hr, Diehl:2003ny}:
\begin{equation}\label{eq:impact-dist}
\mathsf{q^\pi}(x,b) = \int \frac{ d \mathbf{q}}{(2\pi)^2} H^\pi(x,\xi=0,t=-\mathbf{q}^2) e^{i\mathbf{q}\,\cdot \, \mathbf{b} },
\end{equation}
where $b$ is the light-front distance from the center of transverse momentum (CoTM). 
Figure~\ref{fig:tomography} shows slices of the distribution with selected $x \in [0.1,0.9]$, as well as the two-dimensional distribution at $x=0.45$.
The impact-parameter--dependent distribution describes the probability density for a parton with momentum fraction $x$ to be found in the transverse plane at distance $b$ from the CoTM.
It provides a snapshot of the pion in the transverse plane and indicates what might be expected from pion tomography.

\section{Summary and Outlook}
\label{sec:summary}

We presented a state-of-the-art high-statistics lattice calculation of the valence-quark GPD of the pion using the LaMET approach with a next-to-next-to-leading order perturbative matching formula.
The calculation was done at lattice spacing 0.09~fm with physical pion masses and boosted pion momentum around 1.7~GeV with four additional nonzero transferred momenta in the Breit frame.
We performed multistate analyses with multiple source-sink separations to remove excited-state contamination.
We obtained the $x$-dependent valence-quark GPD by fitting the hybrid-scheme--renormalized matrix elements to a phenomenology-inspired functional form.  
We compared our results with prior lattice studies in special limits of the pion GPD, such as the pion PDFs ($Q^2=0$ limit) and pion form factors (forward limit).
This is the first pion valence-quark GPD directly calculated at the physical quark masses.
We also made predictions for higher moments of the generalized form factors that have not yet been calculated and show $x$-dependent tomography of the pion for the first time using lattice QCD.
Our result provides the most reliable lattice prediction of the valence-quark pion GPD to date and will guide upcoming measurements at JLab and the EIC.
Future work will extend the calculation to additional boost momenta and finer lattice spacings to further constrain the systematics in the lattice calculations.

\section*{Acknowledgments}

We thank MILC Collaboration for sharing the lattices used to perform this study. The LQCD calculations were performed using the Chroma software suite~\cite{Edwards:2004sx}.
This research used resources of the National Energy Research Scientific Computing Center, a DOE Office of Science User Facility supported by the Office of Science of the U.S. Department of Energy under Contract No. DE-AC02-05CH11231 through ERCAP;
facilities of the USQCD Collaboration are funded by the Office of Science of the U.S. Department of Energy,
and the Extreme Science and Engineering Discovery Environment (XSEDE), which was supported by National Science Foundation Grant No. PHY-1548562.
The work of HL is partially supported by the US National Science Foundation under grant PHY 1653405 ``CAREER: Constraining Parton Distribution Functions for New-Physics Searches'', grant PHY 2209424,  and by the Research Corporation for Science Advancement through the Cottrell Scholar Award.



\begin{thebibliography}{100}

\bibitem{Achenbach:2023pba}
P.~Achenbach et~al.
\newblock {The Present and Future of QCD}.
\newblock 3 2023.

\bibitem{Arrington:2021biu}
J.~Arrington et~al.
\newblock {Revealing the structure of light pseudoscalar mesons at the
  electron\textendash{}ion collider}.
\newblock {\em J. Phys. G}, 48(7):075106, 2021.

\bibitem{Roberts:2021nhw}
Craig~D. Roberts, David~G. Richards, Tanja Horn, and Lei Chang.
\newblock {Insights into the emergence of mass from studies of pion and kaon
  structure}.
\newblock {\em Prog. Part. Nucl. Phys.}, 120:103883, 2021.

\bibitem{Aguilar:2019teb}
Arlene~C. Aguilar et~al.
\newblock {Pion and Kaon Structure at the Electron-Ion Collider}.
\newblock {\em Eur. Phys. J. A}, 55(10):190, 2019.

\bibitem{Ji:1996nm}
Xiang-Dong Ji.
\newblock {Deeply virtual Compton scattering}.
\newblock {\em Phys. Rev. D}, 55:7114--7125, 1997.

\bibitem{Radyushkin:1997ki}
A.~V. Radyushkin.
\newblock {Nonforward parton distributions}.
\newblock {\em Phys. Rev. D}, 56:5524--5557, 1997.

\bibitem{Diehl:2003ny}
M.~Diehl.
\newblock {Generalized parton distributions}.
\newblock {\em Phys. Rept.}, 388:41--277, 2003.

\bibitem{Belitsky:2005qn}
A.~V. Belitsky and A.~V. Radyushkin.
\newblock {Unraveling hadron structure with generalized parton distributions}.
\newblock {\em Phys. Rept.}, 418:1--387, 2005.

\bibitem{Kumericki:2016ehc}
Kresimir Kumericki, Simonetta Liuti, and Herve Moutarde.
\newblock {GPD phenomenology and DVCS fitting}: {Entering the high-precision
  era}.
\newblock {\em Eur. Phys. J. A}, 52(6):157, 2016.

\bibitem{Burkert:2022hjz}
V.~D. Burkert et~al.
\newblock {Precision studies of QCD in the low energy domain of the EIC}.
\newblock {\em Prog. Part. Nucl. Phys.}, 131:104032, 2023.

\bibitem{Burkardt:2000za}
Matthias Burkardt.
\newblock {Impact parameter dependent parton distributions and off forward
  parton distributions for zeta ---\ensuremath{>} 0}.
\newblock {\em Phys. Rev. D}, 62:071503, 2000.
\newblock [Erratum: Phys.Rev.D 66, 119903 (2002)].

\bibitem{NA7:1986vav}
S.~R. Amendolia et~al.
\newblock {A Measurement of the Space - Like Pion Electromagnetic Form-Factor}.
\newblock {\em Nucl. Phys. B}, 277:168, 1986.

\bibitem{Conway:1989fs}
J.~S. Conway et~al.
\newblock {Experimental Study of Muon Pairs Produced by 252-GeV Pions on
  Tungsten}.
\newblock {\em Phys. Rev. D}, 39:92--122, 1989.

\bibitem{NA10:1987azm}
P.~Bordalo et~al.
\newblock {Observation of a Nuclear Dependence of the Transverse Momentum
  Distribution of Massive Muon Pairs Produced in Hadronic Collisions}.
\newblock {\em Phys. Lett. B}, 193:373, 1987.

\bibitem{NA10:1985ibr}
B.~Betev et~al.
\newblock {Differential Cross-section of High Mass Muon Pairs Produced by a
  194-{GeV}/$c \pi^-$ Beam on a Tungsten Target}.
\newblock {\em Z. Phys. C}, 28:9, 1985.

\bibitem{Barry:2018ort}
P.~C. Barry, N.~Sato, W.~Melnitchouk, and Chueng-Ryong Ji.
\newblock {First Monte Carlo Global QCD Analysis of Pion Parton Distributions}.
\newblock {\em Phys. Rev. Lett.}, 121(15):152001, 2018.

\bibitem{Novikov:2020snp}
Ivan Novikov et~al.
\newblock {Parton Distribution Functions of the Charged Pion Within The xFitter
  Framework}.
\newblock {\em Phys. Rev. D}, 102(1):014040, 2020.

\bibitem{Barry:2021osv}
P.~C. Barry, Chueng-Ryong Ji, N.~Sato, and W.~Melnitchouk.
\newblock {Global QCD Analysis of Pion Parton Distributions with Threshold
  Resummation}.
\newblock {\em Phys. Rev. Lett.}, 127(23):232001, 2021.

\bibitem{Dudek:2012vr}
Jozef Dudek et~al.
\newblock {Physics Opportunities with the 12 GeV Upgrade at Jefferson Lab}.
\newblock {\em Eur. Phys. J. A}, 48:187, 2012.

\bibitem{Accardi:2012qut}
A.~Accardi et~al.
\newblock {Electron Ion Collider: The Next QCD Frontier}: {Understanding the
  glue that binds us all}.
\newblock {\em Eur. Phys. J. A}, 52(9):268, 2016.

\bibitem{AbdulKhalek:2021gbh}
R.~Abdul~Khalek et~al.
\newblock {Science Requirements and Detector Concepts for the Electron-Ion
  Collider}: {EIC Yellow Report}.
\newblock {\em Nucl. Phys. A}, 1026:122447, 2022.

\bibitem{Anderle:2021wcy}
Daniele~P. Anderle et~al.
\newblock {Electron-ion collider in China}.
\newblock {\em Front. Phys. (Beijing)}, 16(6):64701, 2021.

\bibitem{Chavez:2021koz}
Jos\'e Manuel~Morgado Ch\'avez, Valerio Bertone, Feliciano De~Soto~Borrero,
  Maxime Defurne, C\'edric Mezrag, Herv\'e Moutarde, Jos\'e
  Rodr\'\i{}guez-Quintero, and Jorge Segovia.
\newblock {Accessing the Pion 3D Structure at US and China Electron-Ion
  Colliders}.
\newblock {\em Phys. Rev. Lett.}, 128(20):202501, 2022.

\bibitem{Ji:2013dva}
Xiangdong Ji.
\newblock {Parton Physics on a Euclidean Lattice}.
\newblock {\em Phys. Rev. Lett.}, 110:262002, 2013.

\bibitem{Ji:2014gla}
Xiangdong Ji.
\newblock {Parton Physics from Large-Momentum Effective Field Theory}.
\newblock {\em Sci. China Phys. Mech. Astron.}, 57:1407--1412, 2014.

\bibitem{Ji:2017rah}
Xiangdong Ji, Jian-Hui Zhang, and Yong Zhao.
\newblock {More On Large-Momentum Effective Theory Approach to Parton Physics}.
\newblock {\em Nucl. Phys. B}, 924:366--376, 2017.

\bibitem{Lin:2013yra}
Huey-Wen Lin.
\newblock {Calculating the $x$ Dependence of Hadron Parton Distribution
  Functions}.
\newblock {\em PoS}, LATTICE2013:293, 2014.

\bibitem{Lin:2014zya}
Huey-Wen Lin, Jiunn-Wei Chen, Saul~D. Cohen, and Xiangdong Ji.
\newblock {Flavor Structure of the Nucleon Sea from Lattice QCD}.
\newblock {\em Phys. Rev. D}, 91:054510, 2015.

\bibitem{Chen:2016utp}
Jiunn-Wei Chen, Saul~D. Cohen, Xiangdong Ji, Huey-Wen Lin, and Jian-Hui Zhang.
\newblock {Nucleon Helicity and Transversity Parton Distributions from Lattice
  QCD}.
\newblock {\em Nucl. Phys. B}, 911:246--273, 2016.

\bibitem{Lin:2017ani}
Huey-Wen Lin, Jiunn-Wei Chen, Tomomi Ishikawa, and Jian-Hui Zhang.
\newblock {Improved parton distribution functions at the physical pion mass}.
\newblock {\em Phys. Rev. D}, 98(5):054504, 2018.

\bibitem{Alexandrou:2015rja}
Constantia Alexandrou, Krzysztof Cichy, Vincent Drach, Elena Garcia-Ramos,
  Kyriakos Hadjiyiannakou, Karl Jansen, Fernanda Steffens, and Christian Wiese.
\newblock {Lattice calculation of parton distributions}.
\newblock {\em Phys. Rev. D}, 92:014502, 2015.

\bibitem{Alexandrou:2016jqi}
Constantia Alexandrou, Krzysztof Cichy, Martha Constantinou, Kyriakos
  Hadjiyiannakou, Karl Jansen, Fernanda Steffens, and Christian Wiese.
\newblock {Updated Lattice Results for Parton Distributions}.
\newblock {\em Phys. Rev. D}, 96(1):014513, 2017.

\bibitem{Alexandrou:2017huk}
Constantia Alexandrou, Krzysztof Cichy, Martha Constantinou, Kyriakos
  Hadjiyiannakou, Karl Jansen, Haralambos Panagopoulos, and Fernanda Steffens.
\newblock {A complete non-perturbative renormalization prescription for
  quasi-PDFs}.
\newblock {\em Nucl. Phys. B}, 923:394--415, 2017.

\bibitem{Chen:2017mzz}
Jiunn-Wei Chen, Tomomi Ishikawa, Luchang Jin, Huey-Wen Lin, Yi-Bo Yang,
  Jian-Hui Zhang, and Yong Zhao.
\newblock {Parton distribution function with nonperturbative renormalization
  from lattice QCD}.
\newblock {\em Phys. Rev. D}, 97(1):014505, 2018.

\bibitem{Zhang:2018diq}
Jian-Hui Zhang, Xiangdong Ji, Andreas Sch\"afer, Wei Wang, and Shuai Zhao.
\newblock {Accessing Gluon Parton Distributions in Large Momentum Effective
  Theory}.
\newblock {\em Phys. Rev. Lett.}, 122(14):142001, 2019.

\bibitem{Alexandrou:2018pbm}
Constantia Alexandrou, Krzysztof Cichy, Martha Constantinou, Karl Jansen,
  Aurora Scapellato, and Fernanda Steffens.
\newblock {Light-Cone Parton Distribution Functions from Lattice QCD}.
\newblock {\em Phys. Rev. Lett.}, 121(11):112001, 2018.

\bibitem{Chen:2018xof}
Jiunn-Wei Chen, Luchang Jin, Huey-Wen Lin, Yu-Sheng Liu, Yi-Bo Yang, Jian-Hui
  Zhang, and Yong Zhao.
\newblock {Lattice Calculation of Parton Distribution Function from LaMET at
  Physical Pion Mass with Large Nucleon Momentum}.
\newblock 3 2018.

\bibitem{Chen:2018fwa}
Jian-Hui Zhang, Jiunn-Wei Chen, Luchang Jin, Huey-Wen Lin, Andreas Sch\"afer,
  and Yong Zhao.
\newblock {First direct lattice-QCD calculation of the $x$-dependence of the
  pion parton distribution function}.
\newblock {\em Phys. Rev. D}, 100(3):034505, 2019.

\bibitem{Alexandrou:2018eet}
Constantia Alexandrou, Krzysztof Cichy, Martha Constantinou, Karl Jansen,
  Aurora Scapellato, and Fernanda Steffens.
\newblock {Transversity parton distribution functions from lattice QCD}.
\newblock {\em Phys. Rev. D}, 98(9):091503, 2018.

\bibitem{Lin:2018qky}
Huey-Wen Lin, Jiunn-Wei Chen, Xiangdong Ji, Luchang Jin, Ruizi Li, Yu-Sheng
  Liu, Yi-Bo Yang, Jian-Hui Zhang, and Yong Zhao.
\newblock {Proton Isovector Helicity Distribution on the Lattice at Physical
  Pion Mass}.
\newblock {\em Phys. Rev. Lett.}, 121(24):242003, 2018.

\bibitem{Fan:2018dxu}
Zhou-You Fan, Yi-Bo Yang, Adam Anthony, Huey-Wen Lin, and Keh-Fei Liu.
\newblock {Gluon Quasi-Parton-Distribution Functions from Lattice QCD}.
\newblock {\em Phys. Rev. Lett.}, 121(24):242001, 2018.

\bibitem{Liu:2018hxv}
Yu-Sheng Liu, Jiunn-Wei Chen, Luchang Jin, Ruizi Li, Huey-Wen Lin, Yi-Bo Yang,
  Jian-Hui Zhang, and Yong Zhao.
\newblock {Nucleon Transversity Distribution at the Physical Pion Mass from
  Lattice QCD}.
\newblock 10 2018.

\bibitem{Wang:2019tgg}
Wei Wang, Jian-Hui Zhang, Shuai Zhao, and Ruilin Zhu.
\newblock {Complete matching for quasidistribution functions in large momentum
  effective theory}.
\newblock {\em Phys. Rev. D}, 100(7):074509, 2019.

\bibitem{Lin:2019ocg}
Huey-Wen Lin and Rui Zhang.
\newblock {Lattice finite-volume dependence of the nucleon parton
  distributions}.
\newblock {\em Phys. Rev. D}, 100(7):074502, 2019.

\bibitem{Chen:2019lcm}
Jiunn-Wei Chen, Huey-Wen Lin, and Jian-Hui Zhang.
\newblock {Pion generalized parton distribution from lattice QCD}.
\newblock {\em Nucl. Phys. B}, 952:114940, 2020.

\bibitem{Lin:2020reh}
Huimei Liu.
\newblock {Frontiers in lattice nucleon structure}.
\newblock {\em Int. J. Mod. Phys. A}, 35(11n12):2030006, 2020.

\bibitem{Chai:2020nxw}
Yahui Chai et~al.
\newblock {Parton distribution functions of $\Delta^+$ on the lattice}.
\newblock {\em Phys. Rev. D}, 102(1):014508, 2020.

\bibitem{Bhattacharya:2020cen}
Shohini Bhattacharya, Krzysztof Cichy, Martha Constantinou, Andreas Metz,
  Aurora Scapellato, and Fernanda Steffens.
\newblock {Insights on proton structure from lattice QCD: The twist-3 parton
  distribution function $g_T(x)$}.
\newblock {\em Phys. Rev. D}, 102(11):111501, 2020.

\bibitem{Lin:2020ssv}
Huey-Wen Lin, Jiunn-Wei Chen, Zhouyou Fan, Jian-Hui Zhang, and Rui Zhang.
\newblock {Valence-Quark Distribution of the Kaon and Pion from Lattice QCD}.
\newblock {\em Phys. Rev. D}, 103(1):014516, 2021.

\bibitem{Zhang:2020dkn}
Rui Zhang, Huey-Wen Lin, and Boram Yoon.
\newblock {Probing nucleon strange and charm distributions with lattice QCD}.
\newblock {\em Phys. Rev. D}, 104(9):094511, 2021.

\bibitem{Li:2020xml}
Zheng-Yang Li, Yan-Qing Ma, and Jian-Wei Qiu.
\newblock {Extraction of Next-to-Next-to-Leading-Order Parton Distribution
  Functions from Lattice QCD Calculations}.
\newblock {\em Phys. Rev. Lett.}, 126(7):072001, 2021.

\bibitem{Fan:2020nzz}
Zhouyou Fan, Xiang Gao, Ruizi Li, Huey-Wen Lin, Nikhil Karthik, Swagato
  Mukherjee, Peter Petreczky, Sergey Syritsyn, Yi-Bo Yang, and Rui Zhang.
\newblock {Isovector parton distribution functions of the proton on a superfine
  lattice}.
\newblock {\em Phys. Rev. D}, 102(7):074504, 2020.

\bibitem{Gao:2020ito}
Xiang Gao, Luchang Jin, Christos Kallidonis, Nikhil Karthik, Swagato Mukherjee,
  Peter Petreczky, Charles Shugert, Sergey Syritsyn, and Yong Zhao.
\newblock {Valence parton distribution of the pion from lattice QCD:
  Approaching the continuum limit}.
\newblock {\em Phys. Rev. D}, 102(9):094513, 2020.

\bibitem{Lin:2020fsj}
Huey-Wen Lin, Jiunn-Wei Chen, and Rui Zhang.
\newblock {Lattice Nucleon Isovector Unpolarized Parton Distribution in the
  Physical-Continuum Limit}.
\newblock 11 2020.

\bibitem{Zhang:2020rsx}
Kuan Zhang, Yuan-Yuan Li, Yi-Kai Huo, Andreas Sch\"afer, Peng Sun, and Yi-Bo
  Yang.
\newblock {RI/MOM renormalization of the parton quasidistribution functions in
  lattice regularization}.
\newblock {\em Phys. Rev. D}, 104(7):074501, 2021.

\bibitem{Alexandrou:2020qtt}
Constantia Alexandrou, Krzysztof Cichy, Martha Constantinou, Jeremy~R. Green,
  Kyriakos Hadjiyiannakou, Karl Jansen, Floriano Manigrasso, Aurora Scapellato,
  and Fernanda Steffens.
\newblock {Lattice continuum-limit study of nucleon quasi-PDFs}.
\newblock {\em Phys. Rev. D}, 103:094512, 2021.

\bibitem{Alexandrou:2020zbe}
Constantia Alexandrou, Krzysztof Cichy, Martha Constantinou, Kyriakos
  Hadjiyiannakou, Karl Jansen, Aurora Scapellato, and Fernanda Steffens.
\newblock {Unpolarized and helicity generalized parton distributions of the
  proton within lattice QCD}.
\newblock {\em Phys. Rev. Lett.}, 125(26):262001, 2020.

\bibitem{Lin:2020rxa}
Huey-Wen Lin.
\newblock {Nucleon Tomography and Generalized Parton Distribution at Physical
  Pion Mass from Lattice QCD}.
\newblock {\em Phys. Rev. Lett.}, 127(18):182001, 2021.

\bibitem{Gao:2021hxl}
Xiang Gao, Kyle Lee, Swagato Mukherjee, Charles Shugert, and Yong Zhao.
\newblock {Origin and resummation of threshold logarithms in the lattice QCD
  calculations of PDFs}.
\newblock {\em Phys. Rev. D}, 103(9):094504, 2021.

\bibitem{Lin:2020rut}
Martha Constantinou et~al.
\newblock {Parton distributions and lattice-QCD calculations: Toward 3D
  structure}.
\newblock {\em Prog. Part. Nucl. Phys.}, 121:103908, 2021.

\bibitem{Lin:2021brq}
Huey-Wen Lin.
\newblock {Nucleon helicity generalized parton distribution at physical pion
  mass from lattice QCD}.
\newblock {\em Phys. Lett. B}, 824:136821, 2022.

\bibitem{Bhattacharya:2022aob}
Shohini Bhattacharya, Krzysztof Cichy, Martha Constantinou, Jack Dodson, Xiang
  Gao, Andreas Metz, Swagato Mukherjee, Aurora Scapellato, Fernanda Steffens,
  and Yong Zhao.
\newblock {Generalized parton distributions from lattice QCD with asymmetric
  momentum transfer: Unpolarized quarks}.
\newblock {\em Phys. Rev. D}, 106(11):114512, 2022.

\bibitem{Aglietti:1998ur}
U.~Aglietti, Marco Ciuchini, G.~Corbo, E.~Franco, G.~Martinelli, and
  L.~Silvestrini.
\newblock {Model independent determination of the light cone wave functions for
  exclusive processes}.
\newblock {\em Phys. Lett. B}, 441:371--375, 1998.

\bibitem{Martinelli:1998hz}
G.~Martinelli.
\newblock {Hadronic weak interactions of light quarks}.
\newblock {\em Nucl. Phys. B Proc. Suppl.}, 73:58--71, 1999.

\bibitem{Dawson:1997ic}
C.~Dawson, G.~Martinelli, G.~C. Rossi, Christopher~T. Sachrajda, Stephen~R.
  Sharpe, M.~Talevi, and M.~Testa.
\newblock {New lattice approaches to the delta I = 1/2 rule}.
\newblock {\em Nucl. Phys. B}, 514:313--335, 1998.

\bibitem{Capitani:1998fe}
S.~Capitani, M.~Gockeler, R.~Horsley, H.~Oelrich, D.~Petters, Paul E.~L. Rakow,
  and G.~Schierholz.
\newblock {Towards a nonperturbative calculation of DIS Wilson coefficients}.
\newblock {\em Nucl. Phys. B Proc. Suppl.}, 73:288--290, 1999.

\bibitem{Capitani:1999fm}
S.~Capitani, M.~Gockeler, R.~Horsley, D.~Petters, D.~Pleiter, Paul E.~L. Rakow,
  and G.~Schierholz.
\newblock {Higher twist corrections to nucleon structure functions from lattice
  QCD}.
\newblock {\em Nucl. Phys. B Proc. Suppl.}, 79:173--175, 1999.

\bibitem{Ji:2001wha}
Xiang-dong Ji and Chul-woo Jung.
\newblock {Studying hadronic structure of the photon in lattice QCD}.
\newblock {\em Phys. Rev. Lett.}, 86:208, 2001.

\bibitem{Detmold:2005gg}
William Detmold and C.~J.~David Lin.
\newblock {Deep-inelastic scattering and the operator product expansion in
  lattice QCD}.
\newblock {\em Phys. Rev. D}, 73:014501, 2006.

\bibitem{Braun:2007wv}
V.~Braun and Dieter M\"uller.
\newblock {Exclusive processes in position space and the pion distribution
  amplitude}.
\newblock {\em Eur. Phys. J. C}, 55:349--361, 2008.

\bibitem{Chambers:2017dov}
A.~J. Chambers, R.~Horsley, Y.~Nakamura, H.~Perlt, P.~E.~L. Rakow,
  G.~Schierholz, A.~Schiller, K.~Somfleth, R.~D. Young, and J.~M. Zanotti.
\newblock {Nucleon Structure Functions from Operator Product Expansion on the
  Lattice}.
\newblock {\em Phys. Rev. Lett.}, 118(24):242001, 2017.

\bibitem{Detmold:2018kwu}
William Detmold, Issaku Kanamori, C.~J.~David Lin, Santanu Mondal, and Yong
  Zhao.
\newblock {Moments of pion distribution amplitude using operator product
  expansion on the lattice}.
\newblock {\em PoS}, LATTICE2018:106, 2018.

\bibitem{QCDSF-UKQCD-CSSM:2020tbz}
A.~Hannaford-Gunn, R.~Horsley, Y.~Nakamura, H.~Perlt, P.~E.~L. Rakow,
  G.~Schierholz, K.~Somfleth, H.~St\"uben, R.~D. Young, and J.~M. Zanotti.
\newblock {Scaling and higher twist in the nucleon Compton amplitude}.
\newblock {\em PoS}, LATTICE2019:278, 2020.

\bibitem{Horsley:2020ltc}
Roger Horsley, Yoshifumi Nakamura, Holger Perlt, Paul E.~L. Rakow, Gerrit
  Schierholz, Kim Somfleth, Ross~D. Young, and James~M. Zanotti.
\newblock {Structure functions from the Compton amplitude}.
\newblock {\em PoS}, LATTICE2019:137, 2020.

\bibitem{Detmold:2021uru}
William Detmold, Anthony~V. Grebe, Issaku Kanamori, C.~J.~David Lin, Robert~J.
  Perry, and Yong Zhao.
\newblock {Parton physics from a heavy-quark operator product expansion:
  Formalism and Wilson coefficients}.
\newblock {\em Phys. Rev. D}, 104(7):074511, 2021.

\bibitem{Liu:1993cv}
Keh-Fei Liu and Shao-Jing Dong.
\newblock {Origin of difference between anti-d and anti-u partons in the
  nucleon}.
\newblock {\em Phys. Rev. Lett.}, 72:1790--1793, 1994.

\bibitem{Liu:1998um}
K.~F. Liu, S.~J. Dong, Terrence Draper, D.~Leinweber, J.~H. Sloan, W.~Wilcox,
  and R.~M. Woloshyn.
\newblock {Valence QCD: Connecting QCD to the quark model}.
\newblock {\em Phys. Rev. D}, 59:112001, 1999.

\bibitem{Liu:1999ak}
Keh-Fei Liu.
\newblock {Parton degrees of freedom from the path integral formalism}.
\newblock {\em Phys. Rev. D}, 62:074501, 2000.

\bibitem{Liu:2016djw}
Keh-Fei Liu.
\newblock {Parton Distribution Function from the Hadronic Tensor on the
  Lattice}.
\newblock {\em PoS}, LATTICE2015:115, 2016.

\bibitem{Liu:2017lpe}
Keh-Fei Liu.
\newblock {Evolution equations for connected and disconnected sea parton
  distributions}.
\newblock {\em Phys. Rev. D}, 96(3):033001, 2017.

\bibitem{Liu:2020okp}
Keh-Fei Liu.
\newblock {PDF in PDFs from Hadronic Tensor and LaMET}.
\newblock {\em Phys. Rev. D}, 102(7):074502, 2020.

\bibitem{Ma:2017pxb}
Yan-Qing Ma and Jian-Wei Qiu.
\newblock {Exploring Partonic Structure of Hadrons Using ab initio Lattice QCD
  Calculations}.
\newblock {\em Phys. Rev. Lett.}, 120(2):022003, 2018.

\bibitem{Bali:2017gfr}
Gunnar~S. Bali et~al.
\newblock {Pion distribution amplitude from Euclidean correlation functions}.
\newblock {\em Eur. Phys. J. C}, 78(3):217, 2018.

\bibitem{Bali:2018spj}
Gunnar~S. Bali, Vladimir~M. Braun, Benjamin Gl\"a\ss{}le, Meinulf G\"ockeler,
  Michael Gruber, Fabian Hutzler, Piotr Korcyl, Andreas Sch\"afer, Philipp
  Wein, and Jian-Hui Zhang.
\newblock {Pion distribution amplitude from Euclidean correlation functions:
  Exploring universality and higher-twist effects}.
\newblock {\em Phys. Rev. D}, 98(9):094507, 2018.

\bibitem{Joo:2020spy}
B\'alint Jo\'o, Joseph Karpie, Kostas Orginos, Anatoly~V. Radyushkin, David~G.
  Richards, and Savvas Zafeiropoulos.
\newblock {Parton Distribution Functions from Ioffe Time Pseudodistributions
  from Lattice Calculations: Approaching the Physical Point}.
\newblock {\em Phys. Rev. Lett.}, 125(23):232003, 2020.

\bibitem{Sufian:2019bol}
Raza~Sabbir Sufian, Joseph Karpie, Colin Egerer, Kostas Orginos, Jian-Wei Qiu,
  and David~G. Richards.
\newblock {Pion Valence Quark Distribution from Matrix Element Calculated in
  Lattice QCD}.
\newblock {\em Phys. Rev. D}, 99(7):074507, 2019.

\bibitem{Sufian:2020vzb}
Raza~Sabbir Sufian, Colin Egerer, Joseph Karpie, Robert~G. Edwards, B\'alint
  Jo\'o, Yan-Qing Ma, Kostas Orginos, Jian-Wei Qiu, and David~G. Richards.
\newblock {Pion Valence Quark Distribution from Current-Current Correlation in
  Lattice QCD}.
\newblock {\em Phys. Rev. D}, 102(5):054508, 2020.

\bibitem{Radyushkin:2017cyf}
A.~V. Radyushkin.
\newblock {Quasi-parton distribution functions, momentum distributions, and
  pseudo-parton distribution functions}.
\newblock {\em Phys. Rev. D}, 96(3):034025, 2017.

\bibitem{Balitsky:2019krf}
Ian Balitsky, Wayne Morris, and Anatoly Radyushkin.
\newblock {Gluon Pseudo-Distributions at Short Distances: Forward Case}.
\newblock {\em Phys. Lett. B}, 808:135621, 2020.

\bibitem{Orginos:2017kos}
Kostas Orginos, Anatoly Radyushkin, Joseph Karpie, and Savvas Zafeiropoulos.
\newblock {Lattice QCD exploration of parton pseudo-distribution functions}.
\newblock {\em Phys. Rev. D}, 96(9):094503, 2017.

\bibitem{Karpie:2017bzm}
Joseph Karpie, Kostas Orginos, Anatoly Radyushkin, and Savvas Zafeiropoulos.
\newblock {Parton distribution functions on the lattice and in the continuum}.
\newblock {\em EPJ Web Conf.}, 175:06032, 2018.

\bibitem{Karpie:2018zaz}
Joseph Karpie, Kostas Orginos, and Savvas Zafeiropoulos.
\newblock {Moments of Ioffe time parton distribution functions from non-local
  matrix elements}.
\newblock {\em JHEP}, 11:178, 2018.

\bibitem{Karpie:2019eiq}
Joseph Karpie, Kostas Orginos, Alexander Rothkopf, and Savvas Zafeiropoulos.
\newblock {Reconstructing parton distribution functions from Ioffe time data:
  from Bayesian methods to Neural Networks}.
\newblock {\em JHEP}, 04:057, 2019.

\bibitem{Joo:2019jct}
B\'alint Jo\'o, Joseph Karpie, Kostas Orginos, Anatoly Radyushkin, David
  Richards, and Savvas Zafeiropoulos.
\newblock {Parton Distribution Functions from Ioffe time pseudo-distributions}.
\newblock {\em JHEP}, 12:081, 2019.

\bibitem{Joo:2019bzr}
B\'alint Jo\'o, Joseph Karpie, Kostas Orginos, Anatoly~V. Radyushkin, David~G.
  Richards, Raza~Sabbir Sufian, and Savvas Zafeiropoulos.
\newblock {Pion valence structure from Ioffe-time parton pseudodistribution
  functions}.
\newblock {\em Phys. Rev. D}, 100(11):114512, 2019.

\bibitem{Radyushkin:2018cvn}
Anatoly Radyushkin.
\newblock {One-loop evolution of parton pseudo-distribution functions on the
  lattice}.
\newblock {\em Phys. Rev. D}, 98(1):014019, 2018.

\bibitem{Zhang:2018ggy}
Jian-Hui Zhang, Jiunn-Wei Chen, and Christopher Monahan.
\newblock {Parton distribution functions from reduced Ioffe-time
  distributions}.
\newblock {\em Phys. Rev. D}, 97(7):074508, 2018.

\bibitem{Izubuchi:2018srq}
Taku Izubuchi, Xiangdong Ji, Luchang Jin, Iain~W. Stewart, and Yong Zhao.
\newblock {Factorization Theorem Relating Euclidean and Light-Cone Parton
  Distributions}.
\newblock {\em Phys. Rev. D}, 98(5):056004, 2018.

\bibitem{Bhat:2020ktg}
Manjunath Bhat, Krzysztof Cichy, Martha Constantinou, and Aurora Scapellato.
\newblock {Flavor nonsinglet parton distribution functions from lattice QCD at
  physical quark masses via the pseudodistribution approach}.
\newblock {\em Phys. Rev. D}, 103(3):034510, 2021.

\bibitem{Fan:2020cpa}
Zhouyou Fan, Rui Zhang, and Huey-Wen Lin.
\newblock {Nucleon gluon distribution function from 2 + 1 + 1-flavor lattice
  QCD}.
\newblock {\em Int. J. Mod. Phys. A}, 36(13):2150080, 2021.

\bibitem{Sufian:2020wcv}
Raza~Sabbir Sufian, Tianbo Liu, and Arpon Paul.
\newblock {Gluon distributions and their applications to Ioffe-time
  distributions}.
\newblock {\em Phys. Rev. D}, 103(3):036007, 2021.

\bibitem{Karthik:2021qwz}
Nikhil Karthik.
\newblock {Quark distribution inside a pion in many-flavor ( 2+1 )-dimensional
  QCD using lattice computations: UV listens to IR}.
\newblock {\em Phys. Rev. D}, 103(7):074512, 2021.

\bibitem{HadStruc:2021wmh}
Tanjib Khan et~al.
\newblock {Unpolarized gluon distribution in the nucleon from lattice quantum
  chromodynamics}.
\newblock {\em Phys. Rev. D}, 104(9):094516, 2021.

\bibitem{Fan:2021bcr}
Zhouyou Fan and Huey-Wen Lin.
\newblock {Gluon parton distribution of the pion from lattice QCD}.
\newblock {\em Phys. Lett. B}, 823:136778, 2021.

\bibitem{HadStruc:2022yaw}
Colin Egerer et~al.
\newblock {Toward the determination of the gluon helicity distribution in the
  nucleon from lattice quantum chromodynamics}.
\newblock {\em Phys. Rev. D}, 106(9):094511, 2022.

\bibitem{Salas-Chavira:2021wui}
Alejandro Salas-Chavira, Zhouyou Fan, and Huey-Wen Lin.
\newblock {First glimpse into the kaon gluon parton distribution using lattice
  QCD}.
\newblock {\em Phys. Rev. D}, 106(9):094510, 2022.

\bibitem{Fan:2022kcb}
Zhouyou Fan, William Good, and Huey-Wen Lin.
\newblock {Gluon Parton Distribution of the Nucleon from 2+1+1-Flavor Lattice
  QCD in the Physical-Continuum Limit}.
\newblock 10 2022.

\bibitem{Karpie:2021pap}
Joseph Karpie, Kostas Orginos, Anatoly Radyushkin, and Savvas Zafeiropoulos.
\newblock {The continuum and leading twist limits of parton distribution
  functions in lattice QCD}.
\newblock {\em JHEP}, 11:024, 2021.

\bibitem{Zhang:2020gaj}
Rui Zhang, Carson Honkala, Huey-Wen Lin, and Jiunn-Wei Chen.
\newblock {Pion and kaon distribution amplitudes in the continuum limit}.
\newblock {\em Phys. Rev. D}, 102(9):094519, 2020.

\bibitem{Gao:2022iex}
Xiang Gao, Andrew~D. Hanlon, Nikhil Karthik, Swagato Mukherjee, Peter
  Petreczky, Philipp Scior, Shuzhe Shi, Sergey Syritsyn, Yong Zhao, and Kai
  Zhou.
\newblock {Continuum-extrapolated NNLO valence PDF of the pion at the physical
  point}.
\newblock {\em Phys. Rev. D}, 106(11):114510, 2022.

\bibitem{Xiong:2013bka}
Xiaonu Xiong, Xiangdong Ji, Jian-Hui Zhang, and Yong Zhao.
\newblock {One-loop matching for parton distributions: Nonsinglet case}.
\newblock {\em Phys. Rev. D}, 90(1):014051, 2014.

\bibitem{Ma:2014jla}
Yan-Qing Ma and Jian-Wei Qiu.
\newblock {Extracting Parton Distribution Functions from Lattice QCD
  Calculations}.
\newblock {\em Phys. Rev. D}, 98(7):074021, 2018.

\bibitem{Ji:2020ect}
Xiangdong Ji, Yu-Sheng Liu, Yizhuang Liu, Jian-Hui Zhang, and Yong Zhao.
\newblock {Large-momentum effective theory}.
\newblock {\em Rev. Mod. Phys.}, 93(3):035005, 2021.

\bibitem{Gao:2021dbh}
Xiang Gao, Andrew~D. Hanlon, Swagato Mukherjee, Peter Petreczky, Philipp Scior,
  Sergey Syritsyn, and Yong Zhao.
\newblock {Lattice QCD Determination of the Bjorken-x Dependence of Parton
  Distribution Functions at Next-to-Next-to-Leading Order}.
\newblock {\em Phys. Rev. Lett.}, 128(14):142003, 2022.

\bibitem{Chen:2020ody}
Long-Bin Chen, Wei Wang, and Ruilin Zhu.
\newblock {Next-to-Next-to-Leading Order Calculation of Quasiparton
  Distribution Functions}.
\newblock {\em Phys. Rev. Lett.}, 126(7):072002, 2021.

\bibitem{Liu:2019urm}
Yu-Sheng Liu, Wei Wang, Ji~Xu, Qi-An Zhang, Jian-Hui Zhang, Shuai Zhao, and
  Yong Zhao.
\newblock {Matching generalized parton quasidistributions in the RI/MOM
  scheme}.
\newblock {\em Phys. Rev. D}, 100(3):034006, 2019.

\bibitem{Liu:2018uuj}
Yu-Sheng Liu et~al.
\newblock {Unpolarized isovector quark distribution function from lattice QCD:
  A systematic analysis of renormalization and matching}.
\newblock {\em Phys. Rev. D}, 101(3):034020, 2020.

\bibitem{Follana:2006rc}
E.~Follana, Q.~Mason, C.~Davies, K.~Hornbostel, G.~P. Lepage, J.~Shigemitsu,
  H.~Trottier, and K.~Wong.
\newblock {Highly improved staggered quarks on the lattice, with applications
  to charm physics}.
\newblock {\em Phys. Rev. D}, 75:054502, 2007.

\bibitem{Bazavov:2012xda}
A.~Bazavov et~al.
\newblock {Lattice QCD Ensembles with Four Flavors of Highly Improved Staggered
  Quarks}.
\newblock {\em Phys. Rev. D}, 87(5):054505, 2013.

\bibitem{Hasenfratz:2001hp}
Anna Hasenfratz and Francesco Knechtli.
\newblock {Flavor symmetry and the static potential with hypercubic blocking}.
\newblock {\em Phys. Rev. D}, 64:034504, 2001.

\bibitem{Mondal:2020cmt}
Santanu Mondal, Rajan Gupta, Sungwoo Park, Boram Yoon, Tanmoy Bhattacharya, and
  Huey-Wen Lin.
\newblock {Moments of nucleon isovector structure functions in $2+1+1$-flavor
  QCD}.
\newblock {\em Phys. Rev. D}, 102(5):054512, 2020.

\bibitem{Park:2020axe}
Sungwoo Park, Tanmoy Bhattacharya, Rajan Gupta, Yong-Chull Jang, Balint Joo,
  Huey-Wen Lin, and Boram Yoon.
\newblock {Nucleon charges and form factors using clover and HISQ ensembles}.
\newblock {\em PoS}, LATTICE2019:136, 2020.

\bibitem{Jang:2019jkn}
Yong-Chull Jang, Rajan Gupta, Huey-Wen Lin, Boram Yoon, and Tanmoy
  Bhattacharya.
\newblock {Nucleon electromagnetic form factors in the continuum limit from (
  2+1+1 )-flavor lattice QCD}.
\newblock {\em Phys. Rev. D}, 101(1):014507, 2020.

\bibitem{Jang:2019vkm}
Yong-Chull Jang, Rajan Gupta, Boram Yoon, and Tanmoy Bhattacharya.
\newblock {Axial Vector Form Factors from Lattice QCD that Satisfy the PCAC
  Relation}.
\newblock {\em Phys. Rev. Lett.}, 124(7):072002, 2020.

\bibitem{Gupta:2018lvp}
Rajan Gupta, Boram Yoon, Tanmoy Bhattacharya, Vincenzo Cirigliano, Yong-Chull
  Jang, and Huey-Wen Lin.
\newblock {Flavor diagonal tensor charges of the nucleon from (2+1+1)-flavor
  lattice QCD}.
\newblock {\em Phys. Rev. D}, 98(9):091501, 2018.

\bibitem{Lin:2018obj}
Huey-Wen Lin, Rajan Gupta, Boram Yoon, Yong-Chull Jang, and Tanmoy
  Bhattacharya.
\newblock {Quark contribution to the proton spin from 2+1+1-flavor lattice
  QCD}.
\newblock {\em Phys. Rev. D}, 98(9):094512, 2018.

\bibitem{Gupta:2018qil}
Rajan Gupta, Yong-Chull Jang, Boram Yoon, Huey-Wen Lin, Vincenzo Cirigliano,
  and Tanmoy Bhattacharya.
\newblock {Isovector Charges of the Nucleon from 2+1+1-flavor Lattice QCD}.
\newblock {\em Phys. Rev. D}, 98:034503, 2018.

\bibitem{Gupta:2017dwj}
Rajan Gupta, Yong-Chull Jang, Huey-Wen Lin, Boram Yoon, and Tanmoy
  Bhattacharya.
\newblock {Axial Vector Form Factors of the Nucleon from Lattice QCD}.
\newblock {\em Phys. Rev. D}, 96(11):114503, 2017.

\bibitem{Rajan:2017lxk}
Rajan Gupta, Yong-Chull Jang, Huey-Wen Lin, Boram Yoon, and Tanmoy
  Bhattacharya.
\newblock {Axial Vector Form Factors of the Nucleon from Lattice QCD}.
\newblock {\em Phys. Rev. D}, 96(11):114503, 2017.

\bibitem{Bhattacharya:2015wna}
Tanmoy Bhattacharya, Vincenzo Cirigliano, Saul Cohen, Rajan Gupta, Anosh
  Joseph, Huey-Wen Lin, and Boram Yoon.
\newblock {Iso-vector and Iso-scalar Tensor Charges of the Nucleon from Lattice
  QCD}.
\newblock {\em Phys. Rev. D}, 92(9):094511, 2015.

\bibitem{Bhattacharya:2015esa}
Tanmoy Bhattacharya, Vincenzo Cirigliano, Rajan Gupta, Huey-Wen Lin, and Boram
  Yoon.
\newblock {Neutron Electric Dipole Moment and Tensor Charges from Lattice QCD}.
\newblock {\em Phys. Rev. Lett.}, 115(21):212002, 2015.

\bibitem{Bhattacharya:2013ehc}
Tanmoy Bhattacharya, Saul~D. Cohen, Rajan Gupta, Anosh Joseph, Huey-Wen Lin,
  and Boram Yoon.
\newblock {Nucleon Charges and Electromagnetic Form Factors from 2+1+1-Flavor
  Lattice QCD}.
\newblock {\em Phys. Rev. D}, 89(9):094502, 2014.

\bibitem{Kronfeld:2019nfb}
Andreas~S. Kronfeld, David~G. Richards, William Detmold, Rajan Gupta, Huey-Wen
  Lin, Keh-Fei Liu, Aaron~S. Meyer, Raza Sufian, and Sergey Syritsyn.
\newblock {Lattice QCD and Neutrino-Nucleus Scattering}.
\newblock {\em Eur. Phys. J. A}, 55(11):196, 2019.

\bibitem{Lin:2017snn}
Huey-Wen Lin et~al.
\newblock {Parton distributions and lattice QCD calculations: a community white
  paper}.
\newblock {\em Prog. Part. Nucl. Phys.}, 100:107--160, 2018.

\bibitem{Lin:2022nnj}
Huey-Wen Lin.
\newblock {Hadron Spectroscopy and Structure from Lattice QCD}.
\newblock {\em Few Body Syst.}, 63(4):65, 2022.

\bibitem{FlavourLatticeAveragingGroup:2019iem}
S.~Aoki et~al.
\newblock {FLAG Review 2019: Flavour Lattice Averaging Group (FLAG)}.
\newblock {\em Eur. Phys. J. C}, 80(2):113, 2020.

\bibitem{FlavourLatticeAveragingGroupFLAG:2021npn}
Y.~Aoki et~al.
\newblock {FLAG Review 2021}.
\newblock {\em Eur. Phys. J. C}, 82(10):869, 2022.

\bibitem{Ji:2020brr}
Xiangdong Ji, Yizhuang Liu, Andreas Sch\"afer, Wei Wang, Yi-Bo Yang, Jian-Hui
  Zhang, and Yong Zhao.
\newblock {A Hybrid Renormalization Scheme for Quasi Light-Front Correlations
  in Large-Momentum Effective Theory}.
\newblock {\em Nucl. Phys. B}, 964:115311, 2021.

\bibitem{Cao:2021aci}
N.~Y. Cao, P.~C. Barry, N.~Sato, and W.~Melnitchouk.
\newblock {Towards the three-dimensional parton structure of the pion:
  Integrating transverse momentum data into global QCD analysis}.
\newblock {\em Phys. Rev. D}, 103(11):114014, 2021.

\bibitem{Izubuchi:2019lyk}
Taku Izubuchi, Luchang Jin, Christos Kallidonis, Nikhil Karthik, Swagato
  Mukherjee, Peter Petreczky, Charles Shugert, and Sergey Syritsyn.
\newblock {Valence parton distribution function of pion from fine lattice}.
\newblock {\em Phys. Rev. D}, 100(3):034516, 2019.

\bibitem{Ji:1998pc}
Xiang-Dong Ji.
\newblock {Off forward parton distributions}.
\newblock {\em J. Phys. G}, 24:1181--1205, 1998.

\bibitem{Hagler:2009ni}
Ph. Hagler.
\newblock {Hadron structure from lattice quantum chromodynamics}.
\newblock {\em Phys. Rept.}, 490:49--175, 2010.

\bibitem{Martinelli:1987bh}
G.~Martinelli and Christopher~T. Sachrajda.
\newblock {A Lattice Calculation of the Pion's Form-Factor and Structure
  Function}.
\newblock {\em Nucl. Phys. B}, 306:865--889, 1988.

\bibitem{Draper:1988bp}
Terrence Draper, R.~M. Woloshyn, Walter Wilcox, and Keh-Fei Liu.
\newblock {The Pion Form-factor in Lattice {QCD}}.
\newblock {\em Nucl. Phys. B}, 318:319--336, 1989.

\bibitem{Bonnet:2004fr}
Frederic D.~R. Bonnet, Robert~G. Edwards, George~Tamminga Fleming, Randy Lewis,
  and David~G. Richards.
\newblock {Lattice computations of the pion form-factor}.
\newblock {\em Phys. Rev. D}, 72:054506, 2005.

\bibitem{Brommel:2006ww}
D.~Br\"ommel et~al.
\newblock {The Pion form-factor from lattice QCD with two dynamical flavours}.
\newblock {\em Eur. Phys. J. C}, 51:335--345, 2007.

\bibitem{Frezzotti:2008dr}
R.~Frezzotti, V.~Lubicz, and S.~Simula.
\newblock {Electromagnetic form factor of the pion from twisted-mass lattice
  QCD at N(f) = 2}.
\newblock {\em Phys. Rev. D}, 79:074506, 2009.

\bibitem{Aoki:2009qn}
S.~Aoki et~al.
\newblock {Pion form factors from two-flavor lattice QCD with exact chiral
  symmetry}.
\newblock {\em Phys. Rev. D}, 80:034508, 2009.

\bibitem{Nguyen:2011ek}
Oanh~Hoang Nguyen, Ken-Ichi Ishikawa, Akira Ukawa, and Naoya Ukita.
\newblock {Electromagnetic form factor of pion from $N_f=2+1$ dynamical flavor
  QCD}.
\newblock {\em JHEP}, 04:122, 2011.

\bibitem{Chambers:2017tuf}
A.~J. Chambers et~al.
\newblock {Electromagnetic form factors at large momenta from lattice QCD}.
\newblock {\em Phys. Rev. D}, 96(11):114509, 2017.

\bibitem{Koponen:2017fvm}
J.~Koponen, A.~C. Zimermmane-Santos, C.~T.~H. Davies, G.~P. Lepage, and A.~T.
  Lytle.
\newblock {Pseudoscalar meson electromagnetic form factor at high $Q^2$ from
  full lattice QCD}.
\newblock {\em Phys. Rev. D}, 96(5):054501, 2017.

\bibitem{Alexandrou:2017blh}
C.~Alexandrou et~al.
\newblock {Pion vector form factor from lattice QCD at the physical point}.
\newblock {\em Phys. Rev. D}, 97(1):014508, 2018.

\bibitem{Feng:2019geu}
Xu~Feng, Yang Fu, and Lu-Chang Jin.
\newblock {Lattice QCD calculation of the pion charge radius using a
  model-independent method}.
\newblock {\em Phys. Rev. D}, 101(5):051502, 2020.

\bibitem{Wang:2020nbf}
Gen Wang, Jian Liang, Terrence Draper, Keh-Fei Liu, and Yi-Bo Yang.
\newblock {Lattice Calculation of Pion Form Factor with Overlap Fermions}.
\newblock {\em Phys. Rev. D}, 104:074502, 2021.

\bibitem{Gao:2021xsm}
Xiang Gao, Nikhil Karthik, Swagato Mukherjee, Peter Petreczky, Sergey Syritsyn,
  and Yong Zhao.
\newblock {Pion form factor and charge radius from lattice QCD at the physical
  point}.
\newblock {\em Phys. Rev. D}, 104(11):114515, 2021.

\bibitem{Koponen:2015tkr}
J.~Koponen, F.~Bursa, C.~T.~H. Davies, R.~J. Dowdall, and G.~P. Lepage.
\newblock {Size of the pion from full lattice QCD with physical u , d , s and c
  quarks}.
\newblock {\em Phys. Rev. D}, 93(5):054503, 2016.

\bibitem{JeffersonLab:2008jve}
G.~M. Huber et~al.
\newblock {Charged pion form-factor between Q**2 = 0.60-GeV**2 and 2.45-GeV**2.
  II. Determination of, and results for, the pion form-factor}.
\newblock {\em Phys. Rev. C}, 78:045203, 2008.

\bibitem{JeffersonLab:2008gyl}
H.~P. Blok et~al.
\newblock {Charged pion form factor between $Q^2$=0.60 and 2.45 GeV$^2$. I.
  Measurements of the cross section for the ${^1}$H($e,e'\pi^+$)$n$ reaction}.
\newblock {\em Phys. Rev. C}, 78:045202, 2008.

\bibitem{Horn:2007ug}
T.~Horn et~al.
\newblock {Scaling study of the pion electroproduction cross sections and the
  pion form factor}.
\newblock {\em Phys. Rev. C}, 78:058201, 2008.

\bibitem{JeffersonLabFpi-2:2006ysh}
T.~Horn et~al.
\newblock {Determination of the Charged Pion Form Factor at Q**2 = 1.60 and
  2.45-(GeV/c)**2}.
\newblock {\em Phys. Rev. Lett.}, 97:192001, 2006.

\bibitem{JeffersonLabFpi:2000nlc}
J.~Volmer et~al.
\newblock {Measurement of the Charged Pion Electromagnetic Form-Factor}.
\newblock {\em Phys. Rev. Lett.}, 86:1713--1716, 2001.

\bibitem{Brommel:2005ee}
D.~Brommel, M.~Diehl, M.~Gockeler, Ph. Hagler, R.~Horsley, D.~Pleiter, Paul
  E.~L. Rakow, A.~Schafer, G.~Schierholz, and J.~M. Zanotti.
\newblock {Structure of the pion from full lattice QCD}.
\newblock {\em PoS}, LAT2005:360, 2006.

\bibitem{Bali:2013gya}
G.~Bali, S.~Collins, B.~Gl\"assle, M.~G\"ockeler, N.~Javadi-Motaghi, J.~Najjar,
  W.~S\"oldner, and A.~Sternbeck.
\newblock {Pion structure from lattice QCD}.
\newblock {\em PoS}, LATTICE2013:447, 2014.

\bibitem{Amendolia:1986wj}
S.~R. Amendolia et~al.
\newblock {A Measurement of the Space - Like Pion Electromagnetic Form-Factor}.
\newblock {\em Nucl. Phys. B}, 277:168, 1986.

\bibitem{Burkardt:2002hr}
Matthias Burkardt.
\newblock {Impact parameter space interpretation for generalized parton
  distributions}.
\newblock {\em Int. J. Mod. Phys. A}, 18:173--208, 2003.

\bibitem{Edwards:2004sx}
Robert~G. Edwards and Balint Joo.
\newblock {The Chroma software system for lattice QCD}.
\newblock {\em Nucl. Phys. B Proc. Suppl.}, 140:832, 2005.

\end{thebibliography}
\end{document}